\documentclass[amsmath,amssymb,twocolumn,nofootinbib]{revtex4-1}

\usepackage[utf8]{inputenc}
\usepackage{times,bbm}
\usepackage{amsfonts}
\usepackage{xcolor}
\usepackage{mathtools}
\usepackage{subfigure}

\usepackage{graphicx}
\usepackage{float}
\usepackage{csquotes}
\graphicspath{{./Figures/}}
\usepackage{listings}

\lstnewenvironment{algorithm}[1][] %defines the algorithm listing environment
{   
    \lstset{ %this is the stype
        mathescape=true,
        escapeinside={/*@}{@*/},
        frame=tb,
        numbers=left, 
        numberstyle=\tiny,
        basicstyle=\scriptsize\ttfamily, 
        keywordstyle=\color{blue}\bfseries,
        keywords={,input, output, return, define, step,argsort,for,in,range,return}
        numbers=left,
        xleftmargin=.04\textwidth,
        #1 % this is to add specific settings to an usage of this environment (for instnce, the caption and referable label)
    }
}
{}

\usepackage{hyperref}
\hypersetup{
	breaklinks,
	colorlinks,
	linkcolor=black,
	citecolor=black,
	urlcolor=black,
}

\usepackage{dsfont}
\newcommand{\id}{\mathds{1}}

\newcommand{\Tr}{\text{Tr}}
\newcommand{\ket}[1]{\left \vert #1 \right \rangle}
\newcommand{\bra}[1]{\left \langle #1 \right \vert}
\newcommand{\ketbra}[2]{\left \vert #1 \right \rangle \! \!\left \langle #2 \right \vert}

\newcommand{\ZZ}{\mathcal{Z}}
\newcommand{\nn}{\mathbb{N}}
\newcommand{\ee}[1]{\mathbb{E}(#1)}
\newcommand{\rr}{\mathbb{R}}
\newcommand{\cc}{\mathbb{C}}
\newcommand{\pai}[1]{\sigma^{#1}_{i}}
\newcommand{\taui}{{\tau}^{z}_{i}}
\newcommand{\dis}{h}

\definecolor{jens}{rgb}{.2,0.7,.9}
\definecolor{cadmiumgreen}{rgb}{0.0, 0.42, 0.24}

\definecolor{marek}{rgb}{.5,.5,.2}

\begin{document}

\title{Construction of exact constants of motion and effective models for many-body localized systems}

\author{M.\ Goihl}
\author{M.\ Gluza}
\author{C.\ Krumnow}
\author{J.\ Eisert}
\affiliation{Dahlem Center for Complex Quantum Systems, Freie Universit{\"a}t Berlin, 14195 Berlin, Germany}

\date{\today}

\begin{abstract}
One of the defining features of many-body localization is the presence of extensively many quasi-local conserved quantities. These constants of motion constitute a corner-stone to an intuitive understanding of much of the phenomenology of many-body localized systems arising from effective Hamiltonians. They may be seen as local magnetization operators smeared out by a quasi-local unitary. However, accurately identifying such constants of motion remains a challenging problem. Current numerical constructions often capture the conserved operators only approximately 
%or trade desirable properties such as exact commutation with the Hamiltonian against each other, 
restricting a conclusive understanding of many-body localization. 
 In this work, we use methods from 
the theory of quantum many-body systems out of equilibrium to establish a new approach for finding a complete set of exact constants of motion which are in addition guaranteed to  represent Pauli-$z$ operators. By this we are able to construct and investigate the proposed effective Hamiltonian using exact diagonalization. Hence, our work provides an important tool expected to further boost inquiries into the breakdown of transport due to quenched disorder. 
\end{abstract}
\maketitle

\section{Introduction}
The question of the precise mechanism of thermalization of closed quantum many-body systems lies at the 
heart of the foundations of quantum statistical mechanics.
For generic systems, one generally expects that the unitary time evolution evolves the system into states 
that can locally be captured by a thermodynamic ensemble using only few parameters such as the total energy or particle number
\cite{1408.5148,Polkovnikov_etal11,christian_review}. 
This expectation will be violated if additional structure is present in the
system that enforces a \emph{local} memory of initial conditions by confining for instance particles to local regions. 
Such a non-thermalizing behavior caused by localization is most famously observed
due to quenched disorder in Anderson insulators \cite{Anderson} and prevails
under the addition of interactions in the form of 
 \emph{many-body localization}
(MBL) as predicted theoretically 
\cite{Basko,Oganesyan,Laflorencie2015} and observed experimentally \cite{Schreiber842,choi2016exploring}.

These systems are expected to feature extensively many \emph{quasi-local constants of
motion} (qLCOMs) which prevent a 
thermodynamic description.
In stark contrast to the Anderson insulator, many-body localized systems feature a slow, \emph{unbounded growth of 
entanglement} 
due to interactions \cite{Prosen_localisation,Pollmann_unbounded}. 
Moreover, all MBL eigenstates are expected to fulfill an {entanglement area law} 
\cite{Bauer,AreaReview,Laflorencie2015,1409.1252},
which delineates them from the eigenstates
of thermalizing systems while making them amenable to \emph{tensor network approaches}
\cite{PhysRevLett.118.017201,PhysRevLett.116.247204,PollmannCirac,PhysRevB.93.245129,PhysRevX.7.021018,1409.1252}.
Due to their special structure, MBL systems constitute candidates for
understanding fundamental aspects of quantum mechanics, microscopic transport properties and interacting systems as their efficient description appears to be in reach.

One of the most successful explanations of the intriguing behavior of MBL
systems has been through a proposed effective Hamiltonian valid in the strong disorder
limit, stated by employing a complete set of qLCOMs \cite{huse2014phenomenology,1305.5554}. 
This description explains the logarithmic entanglement growth 
\cite{PhysRevLett.110.260601,Prosen_localisation,PhysRevLett.110.260601,1412.5605}.
For the case of disordered spin chains the qLCOMs are considered to be dressed
local magnetization operators, i.e., local spin operators conjugated by a
unitary transformation smearing their support within an exponential envelope
but at the same time promoting them to constants of motion.
Under reasonable assumptions these operators can actually be calculated
analytically for a specific MBL model \cite{Imbrie2016} 
that is disordered in all parameters.
For models which
contain disorder only in form of local potentials, much in the spirit of
current experimental investigations, no analytical results are known as of today.
It is hence unclear if more physical models of MBL, such as the disordered Heisenberg chain, can actually be mapped to the effective Hamiltonian of Refs.~\cite{huse2014phenomenology,1305.5554}. 
In fact, it seems fair to say that a satisfactory machinery to
numerically construct exact quasi-local constants of motion 
%that does not trade wanted features unfavorably
is still lacking. This is what we report progress on.

Among the strategies established so far are several variants of transformation schemes which focus on 
decoupling the Hamiltonian
\cite{PhysRevLett.110.067204,PhysRevLett.116.010404,RademakerNew,monthus2016flow,GilMBL}
and by this implicitly define qLCOMs. These approaches have the advantage of being able to treat larger systems 
at the cost of making specific approximations, whose exact effects need to be 
understood \cite{PhysRevB.93.014203,luitz2017small}.
For small systems exact diagonalization based methods can be used 
%and the properties of explicit conserved quantities can be studied 
\cite{PhysRevB.91.085425,PhysRevB.94.144208,he2016interaction}. 
While in general quite arbitrary operators qualify as constants of motion, one aims to ensure specific attributes, when constructing them numerically. The
qLCOMs are supposed to be quasi-local, resemble Pauli-$z$ operators by being traceless with only two degenerate eigenvalues, and mutually
commute among each other and of course with the Hamiltonian of the system. Different numerical schemes trade these properties differently against each other.
Whenever exact diagonalization is feasible then qLCOMs can be constructed
directly, e.g., via optimizing the commutant matrix
\cite{PhysRevB.94.144208,he2016interaction} or performing the infinite-time
average \cite{PhysRevB.91.085425} which also inspired our work. 
The latter methods perturb the spectrum and the qLCOMs are not dressed spins 
anymore and the former study shows that when a local region is embedded into a larger one then the optimal qLCOMs conditioned on the subsystem size could be a superposition of several dressed spin operators because of tail cancellation.
In neither of these ED studies it was possible to construct the effective Hamiltonian of Ref.\ \cite{huse2014phenomenology} in order to support the large-scale transformation schemes. 

In this work we present a novel scheme for computing constants of motion that allows to study the effective Hamiltonian.
%Based on an exact diagonalization of the system, we construct exact mutually commuting constants of motion with a guaranteed Pauli spectrum as expected for dressed spin-operators. We verify their locality in the MBL phase numerically. 
The idea behind our construction follows a clear physical intuition: Quasi-local conservation of the local magnetization implies that the corresponding local Pauli-$z$ operators remain approximately local under time evolution.
%and hence also in equilibrium which for that reason would be non-thermal.
%In agreement with this argument, it was found in Ref.\ \cite{PhysRevB.91.085425} that in MBL systems infinite time averages of local operators are approximately local.
%Moreover the locality of interactions \cite{arad2016connecting} and pinning to the large on-site disorder \cite{bravyi2011schrieffer} could be possible explanations to our numerical observation that the spectrum of these operators -- upon appropriate energy-level ordering -- roughly resemble Pauli-$z$ operators translating the real space to the energy-eigenstates. 
We show numerically that infinite time averaged magnetization operators can be
promoted to true Pauli-$z$ operators, while keeping the locality properties
intact and gaining the desired spectrum by construction. Our construction fails
to be local if the time evolution ergodically spreads local excitations and is
hence physically directly connected to the breakdown of MBL.
Equipped with a full set of exact qLCOMs we go a step further and study the effective model of Ref.\ \cite{huse2014phenomenology} for the disordered Heisenberg chain. Due to their approximative nature, such a discussion of the effective model based on obtained qLCOMs was missing in previous approaches despite being one of the major practical reasons to find qLCOMs in the first place.

\section{Setting}
We consider the prototypical model of MBL, the disordered spin-$1/2$-Heisenberg chain  
on $L$ sites
\begin{equation}
\label{eq:Heis}     
  H = \sum_{i=1}^L \left(\pai{x}\sigma^{x}_{i+1}+ \pai{y}\sigma^{y}_{i+1}+ \pai{z} \sigma^{z}_{i+1} +\Delta\, \dis_i \pai{z}\right)\,,
\end{equation}
where the $\dis_i$ are drawn from the interval $\dis_i \in [-1,1]$ and $\Delta$ denotes the disorder strength. 
This model is expected to undergo a localization transition at $\Delta \approx 7.5$.
Moreover, we use periodic boundary conditions in order
to have a meaningful definition of support for all lattice sites 
and denote with $\text{dist}(\cdot,\cdot)$ the natural distance of two sites for a ring configuration. %, 
%The Hamiltonian is assumed to contain a \emph{translation-invariant, interacting part} $H_B$ supporting ballistic or diffusive transport and 
%a \emph{disordered part} $H_{R}$ which in the context of MBL is commonly a diagonal random on-site magnetic field.
%In order to study phenomena connected to breakdown of transport it is
%furthermore both useful and physically plausible to impose that $H$ conserves the total magnetization $M$, i.e., satisfy $[H,M]=0$ such that the number of magnetic excitations is a global constant of motion.
The Pauli operators above in the Hamiltonian denote real space spin operators acting on lattice sites $i=1,\ldots,L$ by $\pai{\alpha} := \id_{2^{i-1}}\otimes\sigma^\alpha\otimes\id_{2^{L-i}}$ where $\sigma^\alpha$ for $\alpha=x,y,z$ denotes 
the spin $1/2$-Pauli matrices and $\id_n$ the identity on $\mathbb C^n$. 
These operators are formulated within the standard real space basis $\{\ket{i_1 \dots i_L}|i_j=0,1\}$ which we abbreviate by $\widetilde{\ket{j}}$ with $j=1,\dots,2^L$ and $\widetilde{\ket{j}} = \ket{(j-1)_2}$ where $x_2$ denotes the binary representation of $x\in \nn$ and we add leading 0's on the left such that $x_2$ has always $L$ bits.
For the following, it is useful to note that
the $\pai{z}$ operators for $i=1,\ldots,L$ can be written as
\begin{equation}
  \pai{z}=\sum_{j=1}^{2^L} (-1)^{\lfloor (j-1)/2^{L-i}\rfloor}\widetilde{\ket{j}}\widetilde{\bra{j}}
\end{equation}
with $\lfloor\,\cdot\,\rfloor$ denoting the floor function.

Similarly, we introduce Pauli-operators
defined in energy space.
Given an eigenbasis $\{\ket{k}\}$ of $H$, we specify 
another set of  Pauli-$z$ operators through the relation
\begin{equation}
\ZZ_i=\sum_{k=1}^{2^L} (-1)^{\lfloor (k-1)/2^{L-i}\rfloor}\ketbra{k}{k}.
\end{equation}
In the infinite disorder limit ($\Delta \rightarrow \infty$), the Hamiltonian
becomes diagonal in the real space basis and hence these operators 
become equal to the $\{\pai{z}\}$ operators. For the general case with finite $\Delta$ however, the $\{\ZZ_i\}$ and $\{\pai{z}\}$ are formulated in different bases and differ from each other.
Written in the given eigenbasis of $H$ it holds then that $\ZZ_i =\id_{2^{i-1}}\otimes\sigma^z\otimes\id_{2^{L-i}}$ 
which  corresponds to a formal  decomposition $\mathcal H=\mathbb C ^{2^L} \simeq \otimes_{i=1}^L \mathbb C^2$
that is implicitly fixed by an arbitrarily chosen  \emph{order} of energy eigenvalues and eigenvectors. 
As this is crucial for the following we emphasis that the $\ZZ_i = \id_{2^{i-1}}\otimes\sigma^z\otimes\id_{2^{L-i}}$ operators are formulated in energy-space, meaning that the $\sigma^z$ operators here are diagonal in energy-space and in principle unrelated to their real space versions. Hence, there are two decompositions of the Hilbert space into $\otimes_{i=1}^L \mathbb C^2$, one in real and the other in energy space. Identifying a decompostion of $\mathcal{H}$ in energy space which preserves locality in real space lies at the heart of the construction of the set of qLCOMs. 

Throughout this work, the MBL constants of motion will be denoted by $\taui$. 
Let us summarize their desired properties.

\emph{i) Independent conserved quantities.}
  The $\{\taui\}$ operators must commute with $H$ and each other
  \begin{equation}
  [H,\taui] = 0 \qquad \text{and}\qquad [\taui,\tau_j^z] =0\qquad \forall i,j\,
\end{equation}
(in fact, they should be functionally independent, i.e., no constant of motion can be expressed as a function of the other).

\emph{ii) Dressed spins.}
The qLCOMs are expected to have a spectrum resembling Pauli-$z$ operators, i.e., there exists
a dressing unitary $U_D^\dagger$ transforming the energy to real space
\begin{equation}
\label{deftau}
  \taui = U_D \ZZ_i U_D^{\dagger}\,.
\end{equation}

\emph{iii) Quasi-locality.}
  For each $i$ let us denote by $S$ a ``buffer'' region of odd cardinality $|S|$, i.e. $S:= \{j: \text{dist}(i,j)\leq (|S|-1)/2\}$.
  Then we demand that the conserved quantities must be quasi-local, meaning each $\taui$ is centered around site $i$ and its local reductions fulfill
   % \begin{equation}
   % \label{decay}
   % 1-\frac{1}{2^{|S|}}\Tr\left[\frac{1}{2^{2|S^C|}}\Tr_{S^C}(\taui)\Tr_{S^C}(\taui)^\dagger\right] \leq f(|S|)
   % %\|\taui - \Gamma_S(\taui) \| \leq f(|S|)\,,
  %\end{equation}
  \begin{equation}
    \label{decay}
    1-\frac{1}{2^{|S|+2|S^C| }} \|  \Tr_{S^C}(\taui) \|_2^2 \leq f(|S|)
    %\|\taui - \Gamma_S(\taui) \| \leq f(|S|)\,,
  \end{equation}
  where $\Tr_{S^C}(\cdot)$ denotes the partial trace over the complement of $S$ and $f:\nn\rightarrow \rr^+$ is a suitably -- presumably exponentially -- decaying function.
Acknowledging that $\|A\|_2^2 = \Tr(A^\dagger A)$, this is exactly the 
quantity measuring locality discussed in Ref.\ \cite{PhysRevB.91.085425}, and it implies the locality
discussed in Ref.\ \cite{ImbrieReview}.
Note that there are several possible definitions for measuring the locality of the qLCOMs. 
It is interesting to see that this notion of quasi-locality based on the Hilbert-Schmidt norm is 
the sense in which it is discussed for integrable models 
\cite{ProsenIntegrable,PhysRevLett.114.140601,PhysRevB.92.195121}.
%We choose to follow Ref.\ \cite{PhysRevB.91.085425} 
%and by this show that our construction indeed has the same locality structure as dephased local magnetization operators. 
%We postpone the discussion of alternative 
%notions of locality and implications of the logarithmic growth of entanglement in MBL systems to future work. 

Note that constructing a set of constants of motions fulfilling only properties \emph{i)} and \emph{ii)} can be easily done for systems which allow for exact diagonalization, as any set of $\{\ZZ_i\}$ operators constructed from any eigenbasis of $H$ will automatically satisfy \emph{i)} and \emph{ii)}. Ensuring \emph{iii)} however is non-trivial in this case and can only be obtained by choosing a correct ordering of eigenvectors of $H$ in the eigenbasis.

\section{Effective description of localization}
Assuming the precise knowledge of the set of qLCOMs, it is possible
to identify an effective Hamiltonian in terms of the $\{\taui\}$ operators as by property \emph{i)} and \emph{ii)} the collection of $\{\taui\}$ and products thereof form a basis for all matrices diagonal in the chosen eigenbasis $\{\ket{k}\}$ of $H$.
%In order to capture a localized system, an additional structure is assumed to hold for an effective formulation of the Hamiltonian using the qLCOMs.
Given a set of qLCOMs the effective model \cite{huse2014phenomenology} takes the form
\begin{equation}
  \label{huse}
  H^{(N_\text{eff})}_\text{eff} =  \sum\limits_{i}\omega^{(1)}_i \taui +\sum_{i,j} \omega^{(2)}_{i,j} \taui \tau_j^z + \dots\,,
\end{equation}
where $\dots$ subsumes terms up to a truncation order $N_\text{eff}$ and
$H^{(L)}_\text{eff}=H$ if the order of the expansion  reaches the system size
$N_\text{eff}=L$.
Let us introduce a subscript $\mu\in \{0,1\}^L$, a binary word of length $L$, which determines the
position of the $\{\taui\}$ operators in the chain and define 
$\tau(\mu) = \prod_{i=1}^L (\taui)^{\mu_i}$. There are $2^L$ many
of these configurations covering all possible combinations
of the $\taui$ operators acting on the chain.
Then, for any 
\begin{equation}
H=\sum_{e=1}^{2^L} E_e \ketbra  ee , 
\end{equation}
we may write the full expansion of Eq.~\eqref{huse} to order $N_\text{eff}=L$ as
\begin{equation}
 H=\sum_{\mu} \omega_\mu \tau(\mu),
\end{equation}
with $\omega_\mu=2^{-L} \Tr[H \tau(\mu)]$. Note that according to Eq.\,\eqref{deftau} the $\omega_\mu$ can be calculated in energy space via
$\omega_\mu=2^{-L} \Tr[\text{diag}(E) \ZZ(\mu)]$ if the energies in $\text{diag}(E)$ are ordered according to the ordering of the eigenbasis constructing the $\{\ZZ_i\}$ operators.
This construction can be interpreted as a \emph{Boolean Fourier transform} of the spectrum $E$ \cite{montanaro,dewolf}. 
For a specific model the weights $\{\omega_\mu\}$ can only be calculated that way if the different $\{\ZZ_i\}$ are orthogonal with respect to the Hilbert-Schmidt
 scalar product, which follows from property \emph{ii)}.
In the localizing case putatively realized by MBL systems, two additional restrictions are expected to hold for the couplings $\{\omega_\mu\}$.

\emph{iv) Convergence.}
	The couplings of different orders are expected to fulfill 
		$\omega_\eta \ll \omega_\xi,$
		whenever 
		\begin{equation}
		\sum_{k=1}^L \eta_k>\sum_{l=1}^L \xi_l.
		\end{equation}
	 This would imply that Eq.~\eqref{huse} is expected to be a 
good approximation of the full Hamiltonian for low 
$N_\text{eff} \ll L$.

   \emph{v) Locality.}
  It is expected that the weights $\{\omega_{\mu}\}$ decay with
  the maximal distance of two $\taui$, $|\omega_\mu| \leq g(d(\mu) )$,
where again $g:\nn\rightarrow \rr^+$ is a suitably decaying function and
$d(\mu):= \max\{\text{dist} (i,j): \mu_i= \mu_j=1\}$.

In later parts of this work, we explicitly construct $H_\text{eff}$ and investigate the validity these two properties using ED.

\begin{figure}
  \includegraphics[width=.45\textwidth]{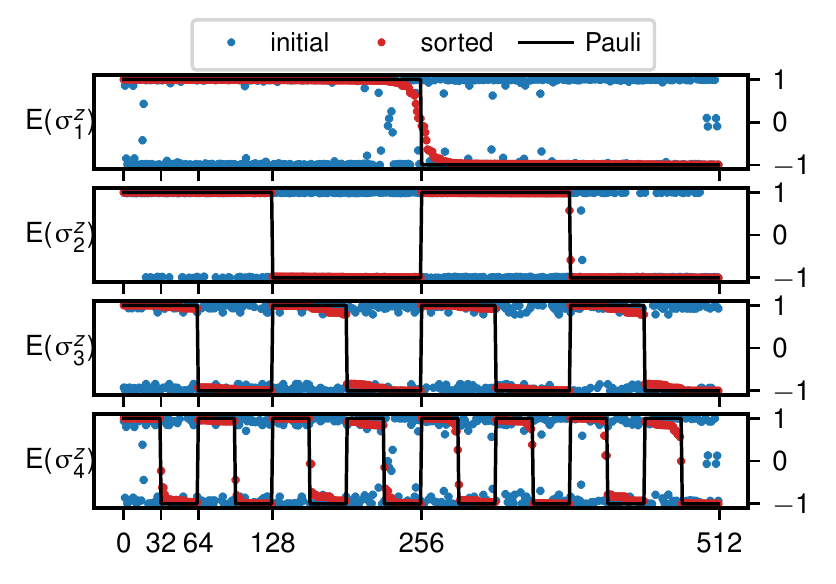}
  \caption{The energy eigenbasis obtained from exact diagonalization (ED) calculations is defined up to a 
  permutation basis which may obscure the physical content available in the
  infinite time average of local spin operators. Here we show data for a
  specific disorder realization with $\Delta = 20$ on $L=9$ sites. The plots show the size of
  the eigenvalues of the infinite time averaged real space Pauli operators
  $\ee{{\pai{z}}}$.
  We illustrate our procedure of permuting the eigenvalues of the set $\ee{{\pai{z}}}$ to obtain
 a particular diagonalization unitary $U_D$ that ensures locality properties of Pauli-$z$ operators $\ZZ_i$ when rotated into the real-space basis $\taui$.
 The difference of $\ee{{\pai{z}}}$ to $\ZZ_i$ comes from the discrepancy of the spectra due to the dephasing. 
 }
  \label{f:proc}
\end{figure}

\section{Mindset of the approach}
The physical intuition behind the algorithm for identifying qLCOMs proposed below is simply that real space spin operators
should merely change under the infinite time average if the system is
localized. Their time average will hence be diagonal in the energy eigenbasis and at the same time quasi-local in real space.
We then set out to find a permutation of the eigenvectors of $H$ such that the time averages of the real space Pauli-$z$ operators
best resemble Pauli-$z$ operators in energy space from which we can then construct the qLCOMs $\{\taui\}$.

The new method to construct the qLCOMs we propose here starts from the energy eigenbasis $\{|e\rangle\}$,
expressed in an arbitrary but fixed ordering.
For each ordering of the eigenbasis 
$|k\rangle = |\pi(e)\rangle, $
where $\pi\in S_{2^L}$ is a {permutation} of the spectrum,
we can define $\{\ZZ_i\}$ 
as above and relate them to real space $\{\taui\}$ operators as in Eq.~\eqref{deftau}. As already pointed out above, these operators by
construction fulfill properties \emph{i)} and \emph{ii)}. Any energy ordering $\pi\in S_{2^L}$ can be used to define a set of $\{\ZZ_i\}$, but 
this in general does not yield quasi-local constants of motion $\{\taui\}$ in real space. 
Demanding property \emph{iii)} in
localized systems, the task is to identify 
permutations $\pi\in S_{2^L}$ that yield local constants of motion.  
However, there are $2^L!$ possible permutations hence achieving global optimality over all permutations
is computationally not feasible. 
Having said that, we can find a solution giving rise to sufficiently
 local constants of motion heuristically, by
exploiting the physical insight above: We order the eigenbasis
such that the spectra of the dephased local magnetization operators 
\emph{simultanously} resemble  Pauli-$z$ spectra of $\{\ZZ_i\}$.
This turns out to be sufficient for ensuring locality of the qLCOMs $\{\taui\}$.

\section{Constructing the set of qLCOMs}
We begin by mapping each real space spin operator $\{\pai{z}\}$ to its infinite time average  
$\ee{\pai{z}} = \sum_{e} \bra{e} \pai{z} \ket{e} \ketbra{e}{e},$
where the sum goes over all eigenvectors $\{\ket{e}\}$ of $H$.
%The name of the operation is
%due to its origin from \emph{equilibration theory}
This operation stems from \emph{equilibration theory} \cite{Linden_etal09,1408.5148} and for 
non-degenerate Hamiltonians one has  
  $\ee{{\pai z}} = \lim_{T\rightarrow\infty} (1/T) \int_0^T \pai{z}(t) \mathrm{d}t$.
This yields $L$ operators diagonal in energy space which commute among each other and with $H$ 
(property \emph{i)}) and are found to be quasi-local \cite{PhysRevB.91.085425} 
(property \emph{iii)}). 
However, due to the non-unitary dephasing, the spectrum of $\ee{\pai{z}}$ does 
not satisfy \emph{ii)} and hence is only approximately Pauli-$z$-like. 
We now set out to reorder the eigenbasis $\{\ket{e}\}$ of each $\{\ee{\pai{z}}\}$ with a permutation $\pi\in S_{2^L}$ such that $\{\ee{\pai{z}}\}$
written in the reordered basis $\{|k = \pi(e)\rangle\}$ best resemble $\{\ZZ_i\}$ in the sense that the entrywise difference between each $\ee{\pai{z}}$ 
and $\ZZ_i$ is small. 
We construct the reordered basis $\{\ket{k}\}$ using a heuristic scheme in multiple steps 
by considering each $\ee{{\pai{z}}}$ successively.

The structure of all $\{\ZZ_i\}$ is by construction known (see for instance the
black dashed line in Fig.~\ref{f:proc} which indicates the diagonal of $\ZZ_1$,
$\ZZ_2$, $\ZZ_3$ and $\ZZ_4$ in the different panels from top to bottom). For
each $\ee{{\pai{z}}}$, we then identify a permutation by which $\ee{{\pai{z}}}$
best approximates $\ZZ_i$ without altering the result identified for previous
$\ee{\sigma^z_j}$ with $j<i$ by sorting the eigenvectors only in the degenerate subspaces of $\ZZ_{i-1}$ according to the size
of the eigenvalues of $\ee{{\pai{z}}}$ and not allowing for a mixing between
those subspaces.
To illustrate the concept consider the operator $\ZZ_1=\sigma^z\otimes \id_{2^{L-1}}$ in the energy eigenbasis. 
It is diagonal in the desired basis $\{\ket{k}\}$ and takes the form $\ZZ_1=\id_{2^{L-1}}\oplus -\id_{2^{L-1}}$. 
Hence, the entrywise closest permutation of $\ee{\sigma^z_1}$ is simply sorting its
spectrum by size (cf. Fig.~\ref{f:proc}, first row). Note that
this choice is highly non-unique, as it allows for an 
arbitrary order inside the two degenerate sectors. We will use
this ambiguity to optimize the remaining qLCOMs.
Next, $\ZZ_2$ has the form $\ZZ_2 = \id_{2^{L-2}}\oplus -\id_{2^{L-2}} \oplus
\id_{2^{L-2}}\oplus -\id_{2^{L-2}}$. The infinite time average
$\ee{\sigma^z_2}$ gives us a new spectrum to optimize.
We then exploit the fact that in the degenerate sectors 
of $\ZZ_1$ our ordering is at the moment arbitrary, i.e., not fixed by
$\ee{{\sigma_1^{z}}}$. In the second step, we therefore sort each of the two
sectors by size of the spectrum of $\ee{{\sigma_2^{z}}}$ (cf. Fig.~\ref{f:proc}, second row).
It is important to note that we must not swap entries from
the different sectors with one another as this would spoil the formerly
established permutations. This procedure is iterated for the remaining
$\ee{{\sigma_i^{z}}}$ as shown in Fig.\,\ref{f:proc}.
Ultimately, for fixing the last permutation $\pi_L\in S_{2^L}$, we only
have the freedom to sort in $2^L/2$ many blocks of size 2, namely to perform swaps for neighboring eigenvectors only. As a result we find the 
final ordering $\pi=\pi_L\circ\dots\circ\pi_1$ and 
we collect the resulting basis to the unitary $U_D$ of Eq.\,\eqref{deftau} 
that can be used to represent the qLCOMs in real space. To be precise,
we now use the obtained $U_D$ to transform the $\{\ZZ_i\}$ which by
construction fulfill properties \emph{i)} and \emph{ii)} into real space.
The following pseudocode describes a possible way to implement this procedure numerically.
We use a notation close to python and denote for instance for a list $l$ of
numbers $1,\dots,N$ in an arbitrary order, a vector $v\in\cc^N$ and a 
matrix $U\in\cc^{N\times N}$ by $v{[l]}$ and $U{[:,l]}$ the vector 
and matrix for which the elements of the vector and columns of the 
matrix are reordered according to $l$, i.e., $v{[l]}_i = v_{l_i}$
and $U{[:,l]}_{i,j} = U_{i,l_j}$. Similarly, we denote for $v\in\cc^{N}$ 
and $1\leq n < m \leq N$ by $v{[n:m]}$ the vector 
$v{[n:m]}\in\cc^{m-n}$ with entries $v{[n:m]}_i = v_{n+i-1}$.
\begin{algorithm}
input: diagonalizing unitary $U$ (as obtained from ED)
input: $L$ real space Pauli operators $\pai{z}$
output: quasi-local diagonalization unitary $U_D$

define infinite_time_average($V$, $O$):
    return $diag(VOV^\dagger)$

perm = $[1,\dots,2^L]$
for $n$ in $\{1,\dots,L\}$:
    spec = infinite_time_average($U$, $\sigma_1^z$)
    permuted_spec = spec[perm]
    for $j$ in $\{1,\ldots,2^n\}$:
    	$\text{perm}[(j-1)*2^{L-n}+1 : j* 2^{L-n}] = $ /*@\\
    	    @*/           /*@$(\text{perm}[(j-1)*2^{L-n}+1 : j* 2^{L-n}])[${\bfseries \color{blue}{argsort}}$($\\
    	    @*/           /*@$\text{permuted\_spec}[(j-1)*2^{L-n}+1\, :\, j* 2^{L-n}])]$@*/

return $U[:,\text{perm}]$
\end{algorithm}

\begin{figure*}
  \includegraphics[width=\textwidth]{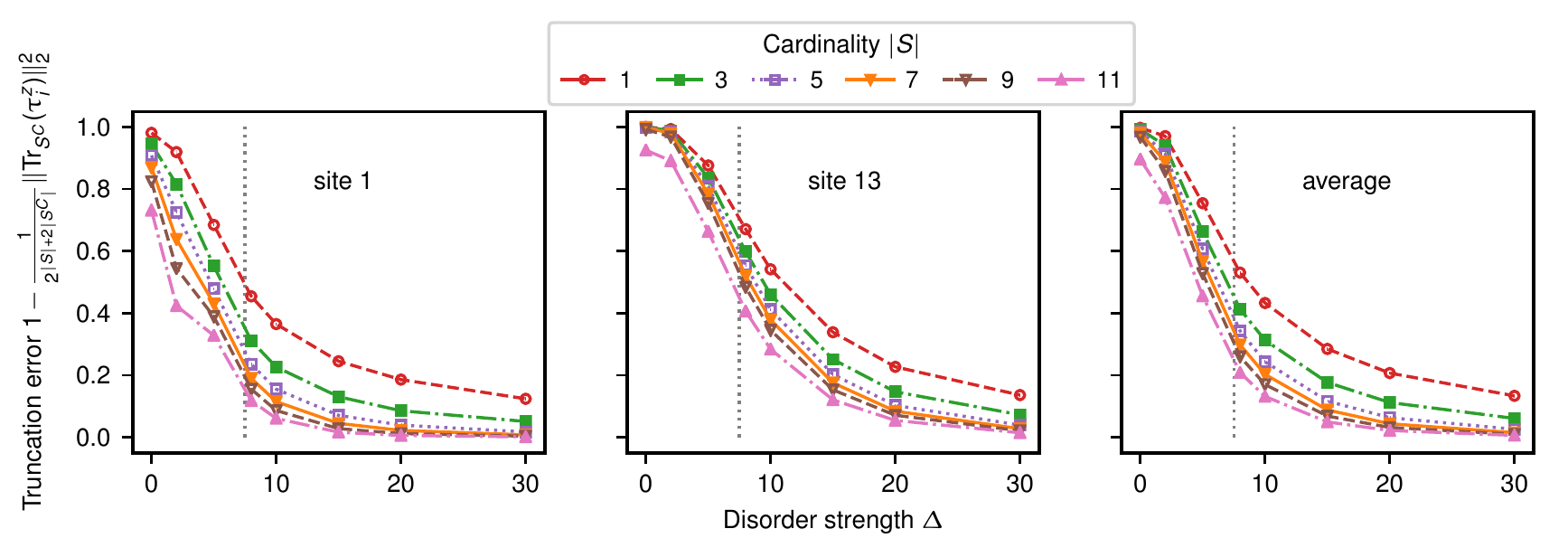}
  \caption{We show the support of the first (left panel) and the last (center) qLCOMs as well as the averaged support of all constructed qLCOMs (right panel) of the disorder Heisenberg chain over the disorder strength $\Delta$. As a measure of support, we use
    the truncation error 
    $1-  \Tr_{S^C}(\taui)\|^2_2/(
    2^{|S|+2|S^C|}
    ) $ for buffer regions $S$ of increasing cardinality $|S|$. 
    Error bars show statistical deviations over 300 realizations.
    We use the Heisenberg Hamiltonian
    on $L =13$ with periodic boundary conditions.
The dashed lines at $\Delta = 7.5$ are a guide to the eye indicating the region of the expected phase transition.}
  \label{f:supp}
\end{figure*}
\section{Numerical results}
\begin{figure}
  \includegraphics[width=0.45\textwidth]{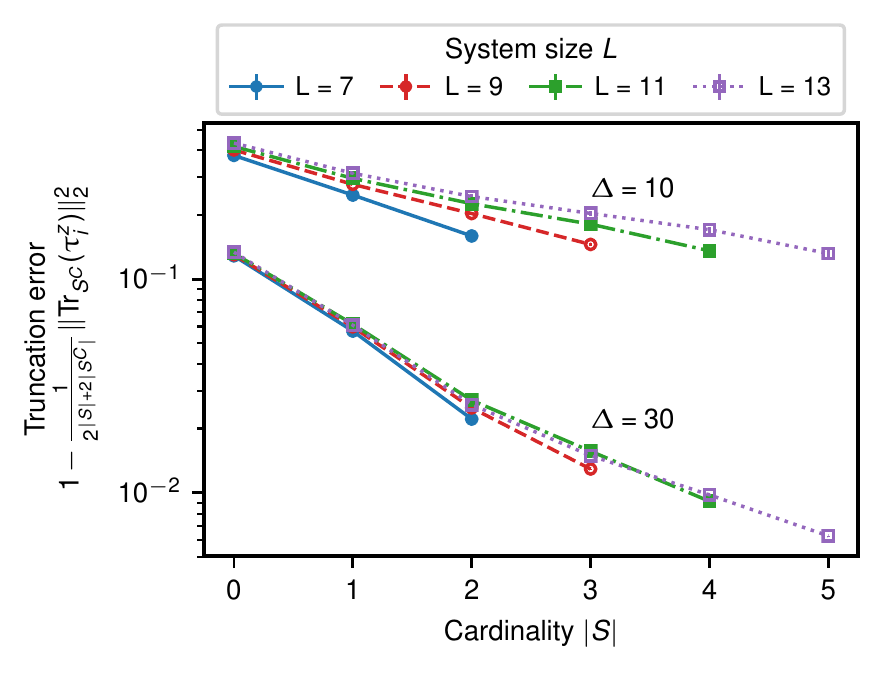}
  \caption{Average truncation error with random disorder for $L \in \{7,9,11,13\}$ with
    $\{1000,1000,1000,300\}$ realizations. 
    We use the Heisenberg Hamiltonian 
    with the disorder strength $\Delta \in \{10,30\}$. Moreover we employ periodic
    boundary conditions. The plot shows the average truncation error
    $1-  \|  \Tr_{S^C}(\taui)\|_2^2/(
    {2^{|S|+2|S^C| }}
    )$
    of the qLCOMs when
    truncated onto a ``buffer'' region of off cardinality $|S|$ averaged over all qLCOMs. 
    The plot is on a $\log$ scale. Lines are guide to the eye.
  Error bars show statistical deviations.}
  \label{tefin}
\end{figure}

\begin{figure}
  \includegraphics[width=0.45\textwidth]{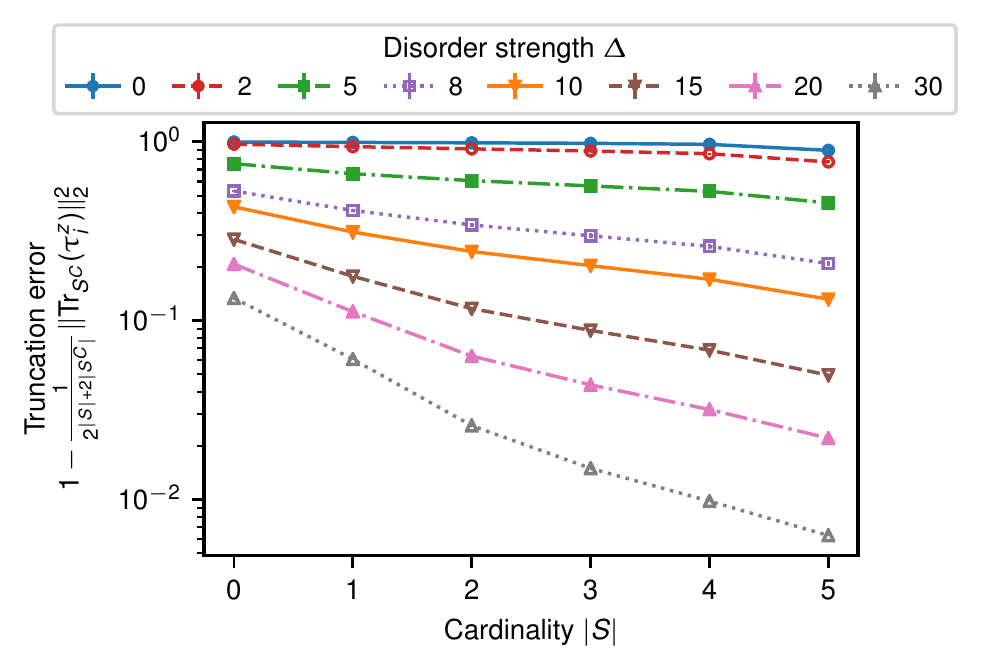}
  \caption{Decay of the average truncation error over all constants of motion
  displayed in Fig. \ref{tefin} on a $\log$ scale for different disorder strengths.
    The error bars indicate the standard deviation of the average and lines are a guide to the eye.
}
  \label{te}
\end{figure}

We now examine the properties of the qLCOMs constructed according to the above
scheme. First, we make sure that the obtained operators are indeed
quasi-local and hence fulfill property \emph{iii)}. 
We find that the qLCOMs constructed with our algorithm
are local to a few sites only at high enough disorder --
an observation, which reproduces the theoretical
predictions.
In Fig.\,\ref{f:supp}  we plot the support of the first, the last and averaged over
all constructed $\{\taui\}$ of the  13-site lattice as a function of disorder
strength $\Delta$ averaged over 300 realizations. 
The quantifier for the support is the truncation error to a subsystem $S$ in 2-norm
defined as 
    $1- \|  \Tr_{S^C}(\taui)\|^2_2 /(2^{|S|+2|S^C|})$. If the value
is close to one, the spectrum of the operator deviates strongly from the Pauli-$z$ spectrum.
If the value is zero, the operator is in this sense 
well-characterized by its reduction to the
subsystem $S$.
We find that
increasing disorder localizes the obtained operators. Additionally, one observes a crossover
in the region of the proposed phase transition.
It can furthermore be seen that despite the recursive nature 
of our approach which allows more variational freedom first
initial qLCOMs
there is only a small systematic error between the first and the last qLCOM 
and all qLCOMs are well 
localized for $\Delta$ large enough. A finite size scaling
is discussed in the following 
indicating that while our method works for the system sizes considered,
it suggests inconclusive results for the locality of the operators for larger systems.
Fig.\ \ref{te} displays the averaged decay of the qLCOMs and shows that in the localized phase 
    $1-  \|  \Tr_{S^C}(\taui)\|^2_2/
    (
    2^{|S|+2|S^C|}
    )$ 
decays exponentially showing that the $\{\taui\}$ are local up to exponential tails.
Here we average both over realizations and qLCOMs per realization. 
Additionally one observes a stronger decay for larger disorder. This scaling with the disorder strength 
is very much expected and consistent with theoretical predictions.
Next, we study the finite size dependence of the locality results. Fig.\,\ref{tefin}
shows the system size dependence of the truncation error 
$1-  \|  \Tr_{S^C}(\taui)\|_2^2/(
    {2^{|S|+2|S^C|}})$
of the qLCOMs for
moderate ($\Delta = 10$) and strong ($\Delta=30$) disorder. The qualitative behavior between
the disorder strengths is consistent with Fig.\,\ref{te}. When considering
increasing system
sizes, we observe that the decay slows down. Nevertheless, we see that
for all system sizes we obtain a strong decay with the distance. 
For the system sizes accessible, we find a still sufficient decay to call the obtained qLCOMs
quasi-local. However, it seems hard to predict the trend for larger systems based on the given data. 
Let us now turn
to insights about the transition between the MBL and the ergodic phase.

\begin{figure}
  \includegraphics[width=0.45\textwidth]{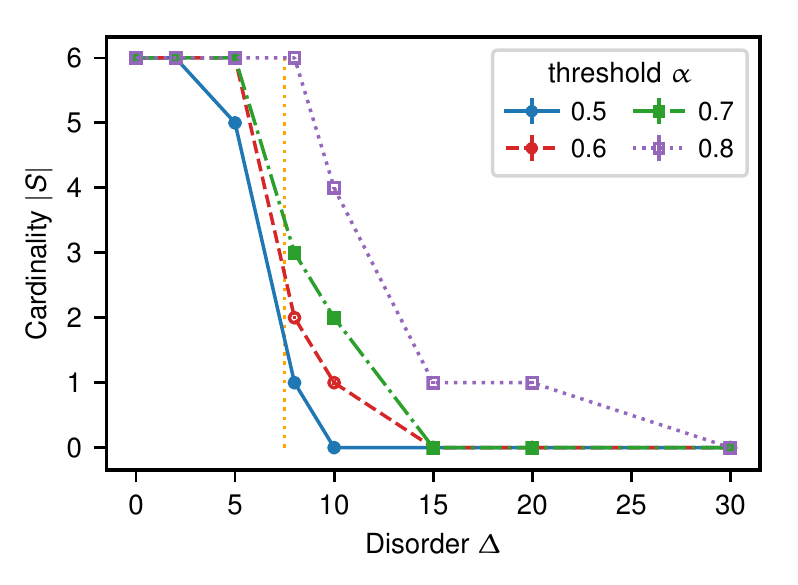}
  \caption{Cardinality of the minimal buffer region 
  for threshold values $\alpha \in \{0.5,0.6,0.7,0.8\}$.
    Values are obtained for the disordered Heisenberg model on
    $L = 13$ with periodic boundary conditions.
    Each data point comprises 300 realizations averaged over all
    qLCOMs.
    Lines are guide to the eye.
    Error bars show statistical deviations.
    The orange dotted line indicates the expected transition at $\Delta = 7.5$.}
  \label{phase}
\end{figure}

An interesting open question is how precisely the picture of the qLCOMs breaks down, once the transition towards
the ergodic phase is being approached. Intuitively, one expects a broadening of the qLCOMs upon
delocalizing, which ultimately leads to completely non-local constants of motion. Here,
we set out to observe this transition in the locality of the calculated qLCOMs.
The measure we employ is the the cardinality of the minimal buffer region $S$
(see above) 
needed to support as much as
a threshold $\alpha$ of the weight of the operator. We again work with the squared two-norm of the reduced
operator as a quantifier of support. We show the results in Fig.~\ref{phase} for different
thresholds $\alpha \in \{0.5,0.6,0.7,0.8\}$. While the resulting curve clearly depends on the chosen
threshold, a transition between a phase, where the operator is supported on the full system
for low disorder and on a single site for high disorder can clearly be observed. To precisely
identify the phase boundary is a challenge for all known methods, and this one is no exception. While 
the measure we propose here may not give a reliable quantitive estimate of the transition, it nevertheless
provides a clear qualitative one. Furthermore, it strengthens the intution of the nature 
of the phase transition, giving rise to a
broadening of qLCOMs.
\begin{figure}
  \includegraphics[width=0.45\textwidth]{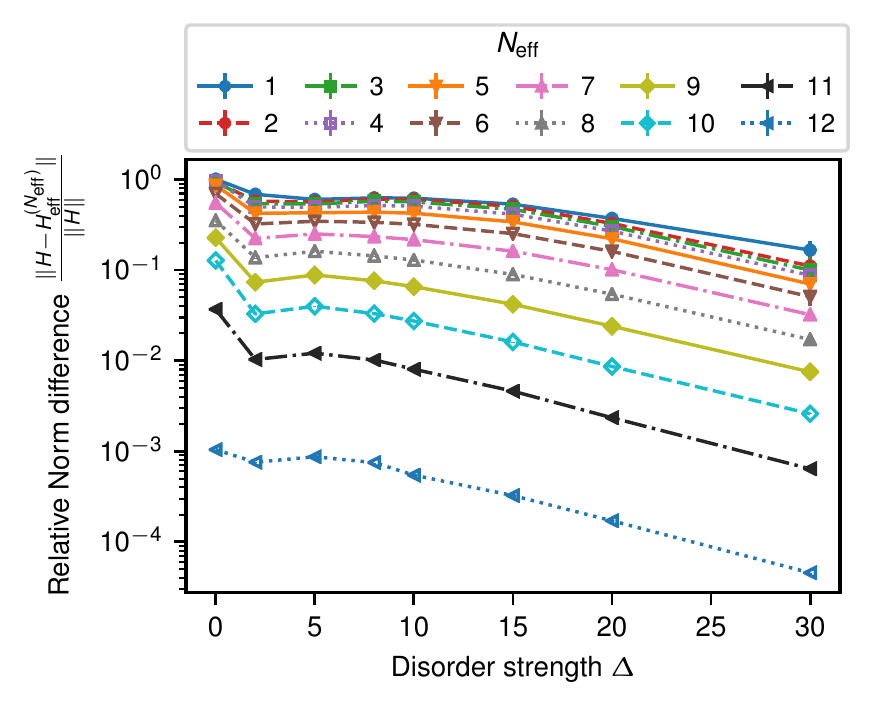}
  \caption{Relative norm difference between effective model and actual Hamiltonian
    ${\|H-H^{(N_\text{eff})}_\text{eff}\|}/{\|H\|}$ on $L = 13$
    with random disorder on a $\log$ scale. Different colors indicate the
    order of the approximation $N_\text{eff}$. 
    Error bars show statistical deviations.
  The average is performed over 300 realizations. Lines are a guide to the eye.} 
  \label{norm}
\end{figure}
\begin{figure}
  \includegraphics[width=0.45\textwidth]{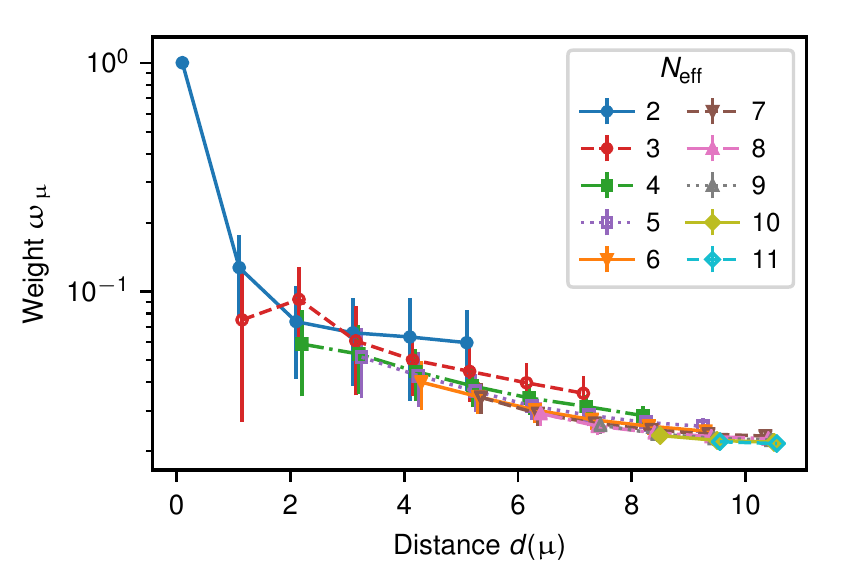}
  \caption{Average coupling strength $\omega_\mu$ on $L = 13$ with random disorder of strength $\Delta = 20$ on a $\log$ scale. 
    Different colors indicate the order of the approximation $N_\text{eff}$.
    Error bars show statistical deviations of the average over 300 
    realizations and per realization over all operators with support of extension $d(\mu)$. 
Lines are a guide to the eye.}
  \label{coup}
\end{figure}
\begin{figure}
  \includegraphics[width=0.9\linewidth]{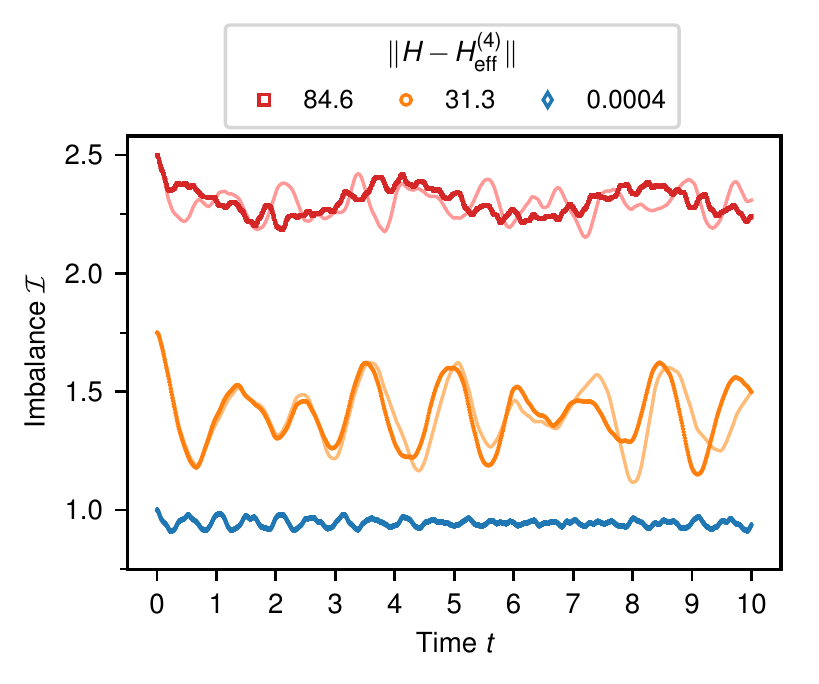}
  \caption{
    Dynamics of the imbalance in the
 Neel state of $L = 13$ spins with periodic boundary conditions with disorder strength $\Delta = 20$.
  The plot shows a comparison of the exact dynamics (solid lines) in three realizations 
(picked by norm difference $\|H-H_\text{eff}^{(4)}\|$ of 30 realizations) with the evolution generated by
the truncated effective Hamiltonian $H_\text{eff}^{(4)}$ (symbols).
  } 
  \label{time}
\end{figure}

Using the constructed qLCOMs, we now turn 
to the effective model and investigate its properties in detail. 
We would like
to point out that this is only possible since our set of qLCOMs fulfills
properties \emph{i)}-\emph{iii)} exactly and not only approximately and hence offers the algebraic structure
necessary to exactly construct the effective description.
We compute the weights $\omega_\mu$ in energy space as explained before using the orthogonality of the $\{\taui\}$ operators and show their decay in Fig.\,\ref{coup}. 
While the $\omega_\mu$ decay with increasing spatial extension 
$d(\cdot)$, there is no apparent inter-order decay. 
Moreover, there is an apparent saturation for higher orders. 
This allows for two explanations. Either the qLCOMs can be further optimized
to fit the expectations of the effective description better or the Heisenberg model cannot be mapped to the effective model
with strongly decaying couplings. 
A possible measure of where to set an effective cut-off $N_\text{eff}$ is the operator norm distance of the Hamiltonian $H$ and its effective description $H^{(N_\text{eff})}_\text{eff}$.
Fig.~\ref{norm} shows the scaling of ${\|H-H^{(N_\text{eff})}_\text{eff}\|}/{\|H\|}$ 
in the dependence on the disorder strength $\Delta$ with $N_\text{eff}$ as a parameter. Here,
we observe that indeed all orders do decay with an exponential trend for larger disorder. However,
to get the norm error small, a rather large $N_\text{eff}$ has to be chosen. 
This seems to put the validity of the effective description as a full solution
in question. However, note that we cannot rule out that qLCOMs can be found
that allow for a better effective model as also stated previously. 
For a brute force approach, $2^L!$ many configurations have to be checked, which
quickly outscales any computational ressources. Hence it will be necessary to work
with a heuristic such as the one presented in this work. Devising new heuristics
which can better fit the effective model 
will be part of future research.

However, imposing such strict global constraints as done by the operator norm difference of the 
exact and effective Hamiltonian may not be required to recover the essential
physical behavior of the system. Hence we will investigate the predictions of the effective model
on a local scale in the following.

\section{Obtaining local dynamics from the effective description} 

In order to 
provide more substance to this discussion, we
investigate the non-equilibrium quench dynamics 
of local observables akin to recent experiments \cite{Schreiber842}.
We compare the dynamical evolution of the
\emph{imbalance} 
\begin{equation}
\mathcal{I} = \frac{1}{L}\sum_i (-1)^i \pai{z} 
\end{equation}
where the initial state is a real space \emph{Neel state vector}
$\ket{1,0,\ldots,1,0,1}$ for
the Heisenberg Hamiltonian and the effective description truncated to order $N_\text{eff}=4$. In Fig.\,\ref{time}
we pick three realizations based on the norm difference $\|H-H^{(4)}_\text{eff}\|$, namely the worst,
intermediately good and best one.
  We find quantitative agreement of the dynamical evolution when the low-order effective description is close in operator 
norm to the true Hamiltonian, however there may exist realizations where the phenomenological model would demand
 many higher-order terms as seen for the bad realization (red in Fig.~\ref{time}).
  Notably, the effective description fails to reproduce quantitatively fast oscillations of the imbalance, 
but the qualitative behavior, e.g., the average imbalance value, is still captured.
  For realizations that work intermediately well, the quantitative agreement is lost over time.

\section{Summary and outlook}

In this work, we have proposed an algorithm for numerically constructing exact constants of motion 
in the localized phase of models exhibiting MBL, with an emphasis on the  random field Heisenberg chain. In contrast to previous attempts of numerically tackling MBL systems, we have put strong emphasis on
exactly fulfilling all desired commutation relations as well as obtaining a Pauli-$z$ spectrum of the constructed operators.
Based on this paradigm, our algorithm finds operators which furthermore act quasi-locally in real space 
in the localized regime. Equipped with a full
set of exact qLCOMs, we are for the first time able to explicitly calculate the effective description of localized systems to all orders. 
It is the hope that this novel tool to construct exact effective Hamiltonians can help to 
satisfactorily explore the rich phenomenology of many-body localized systems.
For future work,
it appears a natural question to investigate whether the equilibrium state of MBL
systems can as anticipated be described by \emph{generalized Gibbs ensembles}
featuring the qLCOM. As MBL systems can be tuned between ''ergodicity'' and
''integrability'', progress in this direction may also shed light on
thermalization in more general models.
Moreover, we aim at elevating the present method to a tensor network consisting of many subsystems, iterating steps, to
give rise to a two-layer quantum cellular automaton, reminiscent of the tensor network of 
Ref.\ \cite{PhysRevX.7.021018}. It is the hope that equipped with exact constants of motion and effective models,
the present work can contribute to resolving the remaining puzzles on many-body localization in one spatial dimension.

%The key challenge to overcome here is to mend the boundaries, which we hope to achieve 
%by Lieb-Robinson based arguments \cite{EisertOsborne06}.

\section{Acknowledgements} 
We thank J.~Behrmann, D.~Litinski and H.~Wilming for helpful comments.
This work has been supported by the ERC (TAQ), the DFG (CRC 183, A2,  EI 519/7-1, and EI 519/14-1),
the Templeton Foundation, and the EC (AQuS). {\it Note added:} After completion of this work, we
became aware of the independent similar work presented in Refs.\ \cite{NewWahl,MBLFlow,mierzejewski2017counting}.

\bibliographystyle{naturemag}
%\bibliography{ConBib}

\end{document}